\documentclass[12pt]{article}

\usepackage{latexsym}
\newcommand{\beq}{\begin{equation}}
\newcommand{\eeq}{\end{equation}}
\newcommand{\ba}{\begin{array}}
\newcommand{\ea}{\end{array}}
\newcommand{\beqa}{\begin{eqnarray}}
\newcommand{\eeqa}{\end{eqnarray}}
\newcommand{\bd}[1]{ \mbox{\boldmath $#1$}  }

\newcommand{\tresj}[6]{\left(\!\!\!
                            \begin{array}{ccc}
                                      #1&#2&#3\\
                                      #4&#5&#6
                            \end{array}\!\!\!
                       \right)}
\newcommand{\seisj}[6]{\left\{\!\!\!
                            \begin{array}{ccc}
                                      #1&#2&#3\\
                                      #4&#5&#6
                            \end{array}\!\!\!
                       \right\}}

\begin{document}
\vskip 3truecm

\begin{center}
{\large {\bf TOROIDAL COLLECTIVE MOTIONS\\ 
IN THE ATOMIC NUCLEUS}} 
\end{center}

\vskip 1truecm

\begin{center}
\c S. Mi\c sicu$^{1,2)}$
\end{center}
\vskip 1truecm

$^{1)}$ National Institute for Nuclear Physics and Engineering,\\ 
~~~~~~Department of Theoretical Physics Bucharest, P.O.Box MG-6, Romania \\

$^{2)}$ Joint Institute for Nuclear Physics, \\
~~~~~~Laboratory for Theoretical Physics, Dubna, Moscow Region, Russia



\begin{abstract}
The work deals with one of the topics of collective motion.
In the frame of {\em Nuclear Fluid Dynamics}, a
model which portrays the nuclear matter as a quantum elastic body, the 
torus-like motions and their associated energies are computed using the 
{\em thirteen moment approximation}. Such excitations correspond to the 
{\em Hill vortex} known from classical Hydrodynamics. There are also 
calculated the nonvanishing contributions of transverse electric form 
factors and differential cross sections in the electroexcitation of these 
collective modes, which are purely toroidal. The spin-dependent collective 
excitations, with toroidal electromagnetic structure, are studied by means 
of the {\em Generalized Goldhaber-Teller} model, with
emphasize on the 1$^{-}$ {\em spin-flip} mode and its excitation in 
spherical nuclei by inelastic electron scattering.
We discuss the importance of toroidal 
contributions in the inclusive electron scattering $(e, e')$ and 
exclusive coincidence electron scattering $(e, e'\gamma)$. In order to 
extract the toroidal multipole, we use the backscattering angles in the 
first mentioned reaction, and the separation of the longitudinal/transverse 
interference in the second case. The introduction of a quantity which 
accounts for the deviations from the {\em Siegert theorem}, shows the 
importance of toroidal quadrupole transitions at high-momentum transfer. 
Another important result concerns the dependence of the intensity of toroidal 
effects on the nuclear {\em vorticity}.
\end{abstract}
\bigskip

\section{Introduction}
The toroidal multipole moments are a distinct family of electromagnetic 
moments, which occur in a special parametrization of charge and curent 
densities \cite{1}, \cite{2}.

Long time ago, Zeldovich showed that a 1/2 spin particle may interact  
with an electromagnetic field not only by means of a dipole electric  
interaction $d(\bd{\sigma}\cdot\bd{E})$ that simultaneously violates  
the space and time inversion, but also by a different type of interaction,    
$a(\bd{\sigma}\cdot\bd{J})$, which violates the parity but not the 
time reversal (Figure 1). 
This new electromagnetic characteristic was 
named {\em anapole} \cite{3} and it is the first member of the class of 
toroidal moments, i.e. the static toroidal dipole moment. Inside nuclei, 
the anapole moment arises as a consequence of $P$-noninvariant \cite{4} 
nuclear forces. 
In classical electrodynamics a very simple example of a toroidal dipole  
is given by a solenoid folded onto a torus. The radiation resistence of the  
corresponding antenna is proportional to
$\left (R_{T}r_{T}^{2}/\lambda^{3}\right )^{2}$, where $r_{T}$ and $R_{T}$ 
are the torus small and large radii \cite{5}. Therefore the toroidal 
antenna depends on the ratio between the geometrical size of the source $d$ 
and the radiation wavelength $\lambda$ like the charge octupole and magnetic  
quadrupole. Such solenoids have non-vanishing magnetic potential in those    
regions of the space where the toroidal dipole moment  $\bd{T}\neq0$
\cite{6}.

The work that we present below discusses the problem of those nuclear  
collective motions associated to purely or partially toroidal 
electromagnetic transitions. There will be also studied the 
properties of excitations associated to vortical nuclear currents  
of orbital isoscalar nature ({\em dipole torus mode} : DTM) 
and of isovector spin dependent nature ({\em spin - isospin mode} : s-is) 
along with the investigation of the possibility to detect them in  
electron inelastic scattering on nuclei.

The electroexcitation of collective rotational and vibrational motions 
with the account of toroidal quadrupole transitions will be also 
investigated. 

\section{The Dipole Torus Mode in Nuclear Fluid Dynamics}
The Nuclear Fluid Dynamics (NFD) allows the description of isoscalar  
giant resonances as in phase harmonic vibrations of proton and neutron 
fluids, the restoring force being a consequence of the elasticity of 
nuclear matter. The nucleus is portrayed as a Fermi fluid which as a 
result of an external perturbation may undergo longitudinal 
oscillations (irotational), as is the case of the liquid drop, but 
also transversal oscillations. Consequently an isoscalar giant resonance 
will be a described by a small amplitude collective oscillation with 
multipolarity $\lambda$, in which protons and neutrons are performing
a divergenceless irotational or vortical displacement \cite{7}.

Basically, the NFD equations can be deduced by taking the classical limit  
of time dependent Hartree-Fock equations (TDHF) 

$$
i\hbar\frac{\partial \hat\rho}{\partial t}=
[~{\hat H},~{\hat \rho}~]
\eqno{(2.1)}\hfill
\label{dfn1}
$$
which is nothing else than the collisionless Boltzmann equation from 
statistical physics of transport phenomena :

$$
\frac{\partial f}{\partial t}+{1\over m}\bd{p}\cdot\nabla_{r}f-
\nabla_{r}U(\bd{r})\cdot\nabla_{p}f=0
\eqno{(2.2)}\hfill
\label{dfn2}
$$
and represents the equation of motion for the Wigner transform of the  
density matrix ${\hat\rho}$
$$
f(\bd{r},\bd{p},t)=\int d \bd{s} 
\rho~(\bd{r}+{\bd{s}\over 2},\bd{r}-{\bd{s}\over 2},t)~
e^{- (i/\hbar) \mbox{\scriptsize{\bf p}} \cdot \mbox{\scriptsize{\bf s}}}
\eqno{(2.3)}
\label{dfn3}
$$
In the argument of $f$ the position vector $\bd{r}$ of a particle is 
determined with a precision larger than the wavelength of the perturbation
in such a way that the Heisenberg principle is not violated.

In order to reduce the complexity of the non-linear partial derivatives 
of equation (2.2), one performs a transformation to coupled equations for 
the macroscopical variables. These physical quantities are introduced as 
$p$-order moments ( = 0 for the density $\rho$, = 1 for the three 
components of the mean velocity $u_{i}(\bd{r},t)$, = 2 for the nine 
components of the strain tensor $P_{ij}(\bd{r},t)$ ) of the distribution
function with respect to the momentum as follows :
$$\rho~(\bd{r},t)=m\int d\bd{p}~f~(\bd{r},\bd{p},t)\hskip 4.cm 
\eqno{(2.4)}\hfill$$
$$\rho~(\bd{r},t)u_{i}(\bd{r},t)=\int d\bd{p}~f~(\bd{r},\bd{p},t)~p_{i}
\hskip 2.77cm\eqno{(2.5)}\hfill$$
$$P_{ij}(\bd{r},t)={1\over m}\int d\bd{p}~f(\bd{r},\bd{p},t)(p_{i}-mu_{i})
(p_{j}-mu_{j})
\eqno{(2.6)}\hfill$$                                        
By integrating the equation of Boltzmann over the momentum $\bd{p}$, with 
weight 1, $p_{j}/m$, $p_{i}p_{j}/m^{2}$, imposing the incompresibility 
condition for the nuclear matter, i.e. $\rho=\rho_{0}=$constant and 
assuming for the strain tensor the ansatz
$$ P_{ij} = P_{0} + p_{ij}\eqno{(2.7)}\hfill$$
one get the linearized NFD equations ({\em the thirteen moments 
approximation})
$$\frac{\partial u_{k}}{\partial x_{k}}=0\hskip 3.6cm\eqno{(2.8)}\hfill$$
$$\rho\frac{\partial u_{i}}{\partial t} + 
\frac{\partial p_{ik}}{\partial x_{k}} = 0\hskip 2.05cm \eqno{(2.9)}\hfill$$
$$\frac{\partial p_{ij}}{\partial t} + P_{0}\left ( 
\frac{\partial u_{i}}{\partial x_{j}} + 
\frac{\partial u_{j}}{\partial x_{i}} \right ) = 0 \eqno{(2.10)}\hfill$$
It is convenient to write the displacement $dx_{i}$ of the nuclear matter 
at a certain point $\bd{r}$ inside the nucleus as
$$
dx_{i} = a_{i}^{\lambda}(\bd{r})\eqno{(2.11)}\hfill$$
or, alternatively by the mean velocity 
$$u_{i} = a_{i}^{\lambda}(\bd{r})\frac{d \alpha_{\lambda}}{d t}
\eqno{(2.12)}\hfill$$
where $a_{i}^{\lambda}(\bd{r})$ is the vector field of instantaneous 
displacements in the Fermi continuum, and $\alpha_{\lambda}$ is the 
time-dependent amplitude of harmonic oscillations 
$(\alpha_{\lambda} \sim \sin\omega_{\lambda}t)$ associated to the resonant 
phenomena that we study in this paper. If we differentiate (2.9) with 
respect to time, (2.10) with respect to   
position $x_{k}$ and we use equation (2.8), we obtain 
$$\rho\frac{\partial^{2} u_{i}}{\partial t^{2}} = 
P_{0}\frac{\partial^{2} u_{i}}{\partial x_{i}^2}\eqno{(2.13)}\hfill$$ 
Afterwards, employing the ansatz (2.12) we arrive at the Helmholz equation
(see (2.1)) for stationary spherical waves
$$\frac{\partial^{2} a_{i}^{\lambda}}{\partial x_{i}^2} + 
k^{2} a_{i}^{\lambda} = 0 \eqno{(2.14)}$$
where $k=\sqrt{\rho\omega^{2}/P_{0}}$ is the wave number. 
Equation (2.14) admits three independent solutions \cite{8}
$$\bd{a}_{l}^{\lambda} = 
N_{l}^{\lambda} \nabla j_{\lambda}(k_{l}r) Y_{\lambda \mu}(\theta,\phi)
\hskip 1.72cm\eqno{(2.15)}\hfill$$
$$\bd{a}_{t}^{\lambda}
= N_{t}^{\lambda} \nabla \times \bd{r} j_{\lambda}(k_{t}r) Y_{\lambda \mu}(\theta,\phi)
\hskip 0.87cm\eqno{(2.16)}\hfill$$
$$\bd{a}_{p}^{\lambda} = 
N_{p}^{\lambda} \nabla \times \nabla \times \bd{r} 
j_{\lambda}(k_{t}r) Y_{\lambda \mu}(\theta,\phi)\eqno{(2.17)}\hfill$$
in a frame with fixed axis. The longitudinal and poloidal solutions 
describe compressional and transversal oscillations of the elastic nuclear 
globe, being responsible for the electric-like resonances with parity  
$\pi = (-)^{\lambda}$, whereas the torsional solution describes 
magnetic-like resonances with parity $\pi = (-)^{\lambda +1}$. 
In the longwavelength limit $(kr\ll 1)$, the poloidal vector field 
becomes proportional to the longitudinal one 
$$
\bd{a}_{p}^{\lambda}(\bd{r})
=N_{p}^{\lambda} \nabla \times \nabla \times \bd{r} r^{\lambda} Y_{\lambda 0}(\theta,\phi)
=N_{p}^{\lambda} (\lambda+1) \nabla r^{\lambda} Y_{\lambda 0}(\theta,\phi) = 
(\lambda +1)\bd{a}_{l}^{\lambda}(\bd{r})
\eqno{(2.18)}\hfill$$
and the torsional field is merely 
$$\bd{a}_{t}^{\lambda}(\bd{r})
=N_{t}^{\lambda} \nabla \times \bd{r} r^{\lambda} Y_{\lambda 0}(\theta,\phi)
\eqno{(2.19)}\hfill$$

It is important to substantiate that in the above mentioned limit, the  
poloidal solution is simultaneously irotational and divergenceless, i.e.
$\nabla \cdot \bd{a}_{p}^{\lambda}~=~
\nabla \times \bd{a}_{p}^{\lambda}~=~0$, whereas the torsional solution 
is purely solenoidal, $\nabla \cdot \bd{a}_{t}^{\lambda}~=0$ 
but $\nabla \times \bd{a}_{t}^{\lambda} \sim \bd{a}_{p}^{\lambda} \neq 0$.

In the dipole case $(\lambda = 1)$, the solution (2.18) corresponds 
to the displacement as a whole of the nucleus, without the change of 
the internal state. The equality (2.18) is obtained from (2.15) and (2.17) 
by keeping the first term in the asymptotic expansion of the spherical 
Bessel function
$$j_{\lambda}(x) \longrightarrow \frac{x^{\lambda}}{(2\lambda+1)!!} 
\left ( 1 - \frac{x^2}{2(2\lambda+3)} + ... \right )\eqno{(2.20)}\hfill$$
Consequently, in order to investigate the dipole response of an 
incompressible elastic globe one need to go beyond the limit imposed 
by the longwavelength approximation and to introduce the high-order terms  
in the expansion (2.20). Then, the poloidal solution becomes 
\cite{9}
$$\bd{a}_{p}^{1} = 
N_{p}^1 \nabla \times \nabla \times \bd{r} r^{3} Y_{10}(\theta,\phi)
\eqno{(2.21)}\hfill$$
In order to establish an explicite expression for the displacements field, 
corrected in such a way to take into account the center of mass motion we 
impose the condition
$$\delta R_{c.m.} = \frac{\int d\bd{r} \rho \bd{a}_{p}^1}{\int d\bd{r} \rho} 
= 0\eqno{(2.22)}\hfill$$
This procedure leads to the following expression for the displacements field 
$$\bd{a}_{p}^{1}  = 
N_{p}^1 \nabla\times\nabla\times\bd{r} r(r^{2}-R^{2})Y_{10}(\theta,\phi)
\hskip 3.2cm$$
$$ =  {2\over \sqrt{3}}N_{p}^{1} 
\left [\sqrt{2}r^{2}\bd{Y}_{12}^{0}(\theta,\phi) +  
(5r^{2}-3R^{2})\bd{Y}_{10}^{0}(\theta,\phi)\right ]  
\eqno{(2.23)}\hfill$$
We can rewrite eq.(2.23) on spherical components
$$(a_{p}^1)_{r} = \sqrt{3\over \pi} N_{p}^{1} (r^{2} - R^{2}) \cos\theta
\hskip 0.53cm \eqno{(2.24)}\hfill$$
$$(a_{p}^1)_{\theta} = -\sqrt{3\over \pi} N_{p}^{1} (2r^{2} - R^{2}) \sin\theta
\eqno{(2.25)}\hfill$$
$$(a_{p}^1)_{\phi} = 0\hskip 4.15cm \eqno{(2.26)}\hfill$$
The dipole poloidal displacement field, or the Dipole Torus Mode (MDT)  
coincide with that for the Hill vortex known from Hydrodynamics 
\cite{10}. 
The Stokes current function corresponding to this vortical flow
is given by 
$$\psi(r,\theta) = N_{p}^{1} (r^{2} - R^{2}) r^{2} \sin^{2}\theta
\eqno{(2.27)}\hfill$$
The contour lines given by eq.(2.27) are ploted in Fig.2. 
The critical or stagnation points are fixed by the conditions
$a_{r} = 0$ and $a_{\theta} = 0$, i.e. 
$r_{c} = R/\sqrt{2}$ \c si $\theta_{c} = \pm\pi$. 
The geometric locus of these points is represented by a ring  
in the equatorial plane of the sphere. The vortical flow of the fluid 
takes place around this ring. The stream lines rotated around the globe 
axis generate tori. Such a vortical flow is known in Classical 
Hydrodynamics under the name of Hill ring vortex, contrary to the linear 
vortex where the critical points are located on the symmetry axis of the  
spheroid. We called the collective excitation corresponding to the Hill   
vortex of the nucleus {\em dipole torus mode} 
( DTM ) \cite{9}. 
A similar kind of collective motion have been studied in \cite{11} 
and \cite{12}.
In the equatorial plane, i.e. the plane containing the vortical ring 
( $\theta = {\pi\over 2}$ )
$$
\bd{\zeta} = \left (\nabla \times \bd{u}_{p}\right )_{\theta=\frac{\pi}{2}} = 
5\sqrt{{3\over \pi}}N_{p}^{1} \dot \alpha (t) r \bd{e}_{\phi}
\eqno{(2.28)}\hfill$$
Thus, the vorticity depends on the radial coordinate.
The energy is given by 
$$E(1^{-}_{tor}) = \sqrt{21\over 5} \hbar \omega_{F} \approx 2\hbar\omega
\eqno{(2.29a)}$$
Therefore, DTM may be interpreted as a dipole transversal isoscalar 
resonance of $2\hbar\omega$ type. Using realistic parameters for the  
Fermi distribution we get \cite{13}, 
$$E(1^{-}_{tor}) = 93.72 A^{-1/3} {\rm MeV}\eqno{(2.29b)}\hfill$$
i.e. the predicted mode is most probably located between the giant 
isovector resonance and the giant isoscalar octupole 
resonance.

We would like to complete the analisys on toroidal dipole by presenting  
the calculation of the transverse electric form factor and of the 
current and transition vorticity densities using the Born's plane wave
approximation. The knowledge of these quantities is important because 
the form facor may be directly measured in electron inelastic scattering
processes \cite{14}.

In the case of the DTM, the current density associated to the transition is 
given in the fluid-dynamic representation by
$$\bd{J}_{tor} = n_{e}\bd{u}_{p} = n_{e}\bd{a}_{p}^{1}(\bd{r})\dot\alpha(t)
\eqno{(2.30)}\hfill$$
where $n_{e} = eZ/An_{0}$ and $n_{0} = 3A/4\pi R^{3}$ is the particle 
density ; $\dot\alpha(t) = \alpha_{0}\omega\cos\omega t$ and
$\alpha_{0} = \left (\frac{\hbar}{2BC} \right )^{1/2}$  
($B_{1}=\frac{6}{7\pi}(N_{p}^{1})^{2}MR^{4}, 
C_{1}=\frac{18}{5\pi}(N_{p}^{1})^{2}Mv_{F}R^{2}$).
We introduce the electric transverse form factor 
$$
| F_{\lambda}^{el}(k) |^{2} = \frac{4\pi}{3}
\langle | \hat{T}_{\lambda 0}^{el}(k,t) |^{2} \rangle _{t}
\eqno{(2..31)}\hfill$$
where by $<...>_{t}$ we understand the time averaging. 
Normalizing to $c^{2}$ one obtains the dimensionless quantity
$$
{1\over c^{2}}| F_{1}^{el}(k) |^{2} = 
\frac{1}{(9\pi)^{1/3}}\sqrt{35\over 6}
\left ( \frac{\omega}{kc}j_{3}(kR) \right )^{2} \frac{Z^2}{A^{4/3}}
\eqno{(2.32)}\hfill$$
In Fig.3 we represented $|F_{1}^{el}(k)|^{2}$ 
for three spherical nuclei : $^{40}$Ca, $^{90}$Zr and $^{208}$Pb. 
This plot exquibite an enhancement of the DTM in heavy nuclei compared  
to light nuclei. Moreover, we notice that the first diffraction maxima 
of the form factor is shifted towards small momentum transfer when we 
pass to nuclei with $Z$ and $A$ large.

A simple calculation shows that the longitudinal and magnetic 
multipoles \cite{2} vanish for the current density (2.30). Thus,   
the transverse electric form factor has the only non-vanishing contribution 
in the excitation with electromagnetic probes (photons,electrons) of DTM!
The Coulomb multipoles does not participate in the excitation due to the 
fact that the charge transition density $\rho_{1}^{tor}$ vanishes as a 
a consequence of the incompresibility condition (2.8) imposed to the  
Fermi globe. In other words, since the excited mode is purely rotational,  
the longitudinal form factor vanishes. Although the magnetic multipole  
should be associated to the excitations of rotational motions, these 
are of oposite parity to DTM, which has a natural parity like the 
electric excitations.

The electroexcitation differential cross-section of DTM looks as follows 
$$
\left ( \frac{d \sigma}{d \Omega} \right )_{1^{-}tor} = 
4\pi\sigma_{Mott} f_{rec}^{-1} 
\left ( \frac{q_{\mu}^{2}}{2\bd{q}^{2}} + \tan^{2}{\theta\over 2} \right )
| \langle 1^{-} \parallel \hat{T}_{1}^{el}(q) \parallel 0^{+} \rangle |^{2}
\eqno{(2.33)}$$
The dependence of (2.33) on the scattering angle is given in 
Fig.4. We should notice that since the longitudinal part does not 
contribute to the electroexcitation of DTM, it does not appear necessary 
to consider the backscattering case. However, in the electroexcitation  
process there will occur also modes which contribute to the longitudinal
part and consequently in order to separate their influence it is better  
to chose the case with scattering angle $\theta =$ 180$^{0}$.

An important characteristic in the study of collective oscillations  
by inelastic electron scattering is given by the transition current 
density , a quantity susceptible to experimental determination.
Basically we are interested in the multipole component 
${\cal J}_{\lambda\lambda +1}(r)$ which can be expressed as inverse 
Fourier transform of the transverse electric form factor \cite{15} : 
$$
{\cal J}_{12}(r) = -\frac{1}{\sqrt{3\pi^{3}}}
\int_{0}^{\infty} F_{1}^{el}(q) j_{2}(qr) q^{2} dq
\eqno{(2.34)}\hfill$$
which integrated gives  
\beqa
{\cal J}_{12}(r)
& = & -\frac{\gamma}{2\sqrt{3\pi}}\frac{r^{2}}{R^{4}},~~~~~~0< r < R
\nonumber\\
& = & -\frac{\gamma}{4\sqrt{3\pi}}\frac{1}{R^{2}},~~~~~~~~r = R
\nonumber\\
& = & ~0,~~~~~~~~~~~~~~~~~~~~r>R
\nonumber
\eeqa
This radial function is ploted in fig.5.

Another quantity of interest in $(e, e')$ inelastic processes is the  
vorticity $\omega_{\lambda\lambda}$. It determines the 
nuclear current properties unconstrained by the charge-current 
conservation law. The vorticity transition density is \cite{16}
$$
\omega_{11}(r) = 
\sqrt{3}\left ( \frac{d}{dr} + \frac{2}{r} \right ) {\cal J}_{12}(r) 
\eqno{(2.35)}\hfill$$
This equation tells us that inside the nucleus the vorticity density 
varies linearly with the radius in the same manner as the vorticity 
vector (2.28). This radial function is represented in Fig.6.

Introducing the multipolar parametrization of Dubovik and Cheshkov 
\cite{2} the transverse electric multipole corresponding to DTM 
reads 
$$\hat{T}_{10}^{el}(q) = q^{2}\hat{T}_{10}^{tor}\eqno{(2.36)}\hfill$$
if we take into account that the Coulomb multipole 
$\hat{Q}_{10}=0$. 
For the toroidal dipole moment associated to the transition 
$0^{+}\rightarrow 1^{-}_{tor}$ the following proportionality
relation is available
$$\left < {\cal T}_{1} \right >_{t} \sim \alpha Z\eqno{(2.37)}\hfill$$
A similar result was known for the electron transition 
$1s_{1/2}\rightarrow 2p_{1/2}$ in Hydrogen-like atoms
\cite{17}. This is the reason why the electromagnetic effects 
of toroidal nature are so weak for small $Z$. 
But whereas in the above mentioned atomic transition the toroidal 
moment enters as a small correction, in the DTM transition it is the   
principal electromagnetic characteristic of the nuclear response.

\section{The spin-flip resonance}
If we consider the excitation of nuclei with $0^{+}$ ground state, then the   
Wigner supermultiplet theory \cite{18} leads to the classification  
of giant dipole resonances given in Table 1. 
In the frame of generalized Goldhaber-Teller mode for isobaric 
nuclei ($N=Z$), having ground state with $J=0^{+}$, $T=0$, the current 
density associated to the GT mode is 
$$\bd{J}(\bd{r}) = {1\over 2}\dot{\bd{q}_{n}} \rho_{0}(\bd{r})
\eqno{(3.1)}\hfill$$
and the spherical components of the magnetization density, for the 
s-is and sw modes, are given by
$$(\mu_{\nu})_{s-is} = \frac{\hbar}{4mc}\frac{g_{p}-g_{n}}{2}\delta_{\nu\nu'}
\bd{q}_{n}\cdot\nabla\rho_{0}(\bd{r})\eqno{(3.2)}\hfill$$
$$(\mu_{\nu})_{sw~~} = \frac{\hbar}{4mc}\frac{g_{p}+g_{n}}{2}\delta_{\nu\nu'}
\bd{q}_{n}\cdot\nabla\rho_{0}(\bd{r})\eqno{(3.3)}\hfill$$
where $\rho_{0}(\bd{r})$ is the ground state density, and $\bd{q}_{n}$ is 
the relative coordinate. The contribution of the sw state (3.3) may be 
neglected due to the smallness of the factor 
$[(g_{p}+g_{n})/(g_{p}-g_{n})]^{2}$.

\begin{table}[h]
\begin{center}
\begin{tabular}{|c|c|c|c|c|}
\hline
\hline
\hskip 2.cm{\tt States}~~~~~~~~~~~~~&~~~~~~$L$~~~~~~&~~~~~~$S$~~~~~~&
~~~~~~~~$J$~~~~~~~~&~~~~~~$T$~~~~~~\\
\hline
\hline
                      &    &   &                         &  \\
Goldhaber - Teller    & 1  & 0 &   1$^{-}$               & 1\\
      (GT)            &    &   &                         &  \\
Spin - Isospin        & 1  & 1 & 0$^{-}$, 1$^{-}$ 2$^{-}$& 1\\
      (s-is)          &    &   &                         &  \\
Spin Wave             & 1  & 1 & 0$^{-}$, 1$^{-}$ 2$^{-}$& 0\\
      (us)            &    &   &                         &  \\
\hline
\end{tabular}
\vspace{3mm}
\label{tabela1}
\end{center}
\end{table}

The calculation of electromagnetic multipoles shows that whereas the GT 
mode is mainly longitudinal the s-is one is purely toroidal like DTM.
It is worthwhile to make a comparison between the electroexcitation 
differential cross-sections of the isovector resonances $1^{-}$ GT and 
$1^{-}$ s-is.
$$\left ( \frac{d \sigma}{d \Omega} \right )_{1^{-}GT~}= 
\sigma_{Mott} b^{2} \frac{F^{2}(q)}{2A} 
\left [ q^{2} V_{L}(\theta) + 2\left ({\omega\over c}\right )^{2}
V_{T}(\theta) \right ]\eqno{(3.4)}\hfill$$
$$\left ( \frac{d \sigma}{d \Omega} \right )_{1^{-}s-is}= 
\sigma_{Mott} b^{2} \frac{F^{2}(q)}{2A}
\left ( \frac{q^{2}\hbar c}{2mc^{2}} \right )
\left ( \frac{g_{p} - g_{n}}{2} \right )^{2} V_{T}(\theta)
\eqno{(3.5)}\hfill$$ 
where $b=\sqrt{m\omega/\hbar}$ is the characteristic length of the  
harmonic oscillator with frequency $\omega$, and the charge density 
form factor in the ground state is 
$$F(q) = \int d\bd{r} e^{i \mbox{\scriptsize{\bf q}} 
\cdot \mbox{\scriptsize{\bf r}}} \rho_{0}(\bd{r})\eqno{(3.6)}\hfill$$

In Fig.7 we have ploted the differential cross-sections of the two 
resonances, GT and s-is, corresponding to the electroexcitation of the 
$^{12}$C and $^{16}$O nuclei. Notice that for small scattering angles, 
the differential cross-sections of the GT resonances are much larger 
than those for the s-is resonances, even at large momentum transfer. 
However, for backscattering angles, the differential cross-sections    
of the s-is modes become important and exceed those of GT resonances 
at momentum transfer $q>0.5$fm$^{-1}$.

\section{Toroidal quadrupole transitions in the Riemann Rotational Model}
The dynamic character of the nuclear rotational motion is one of the basic 
problems, unsolved yet, in nuclear structure theory. 
Many attempts have been made up to the present time to clarify whether  
the nuclear matter can be portrayed as a quantum fluid that could 
support only irotational flows (IF) or it is a quantum rotor 
which gives rise to a rigid rotation (RR) of the whole nucleus.
In a naive image irotational flow may be viewed as a deformation    
which propagates on the surface of the nucleus, along with a 
corresponding motion of the intrinsic structure with small angular
momentum (Fig.8c). At the other extreme lays the rigid body rotation 
(Fig.8a). 

The calculation of inertia moments of these two different types of flow
underestimates the predictions of IF and overestimates those of RR, in 
comparison with the experimental values.
Thence one must consider a model which takes into account the existence 
of currents with intermediate values between IF and RR.

The Riemann rotational model \cite{19} is a simple generalization
of the Bohr-Mottelson model, with the current ranging between the  
limits mentioned above. Moreover the associated velocity field is supposed  
to depend linearly on position. 
A Riemann rotator is an elipsoid whose principal axes have lengths 
$a_{1}, a_{2}, a_{3}$ and in stationary conditions are at rest with 
respect to a frame rotating with constant angular velocity $\bd{\omega}$.
In this rotating frame internal motions with vorticity 
$\bd{\zeta}=\nabla\times\bd{u}$ occur. These two vectors are parallel  
and oriented along one of the ellipsoid principal axes.
Thence the velocity field measured by an observatory at rest with  
respect to the rotating frame is given by
$$
u_{1} = -\frac{a_{1}^{2}}{a_{1}^{2}+a_{2}^{2}}\zeta_{3}x_{2} + 
\frac{a_{1}^{2}}{a_{1}^{2}+a_{3}^{2}}\zeta_{2}x_{3}\eqno{(4.1a)}\hfill$$
$$
u_{2} = -\frac{a_{2}^{2}}{a_{2}^{2}+a_{3}^{2}}\zeta_{1}x_{3} + 
\frac{a_{2}^{2}}{a_{2}^{2}+a_{1}^{2}}\zeta_{3}x_{1}\eqno{(4.1b)}\hfill$$
$$
u_{3} = -\frac{a_{3}^{2}}{a_{1}^{2}+a_{3}^{2}}\zeta_{2}x_{1} + 
\frac{a_{3}^{2}}{a_{3}^{2}+a_{2}^{2}}\zeta_{1}x_{2}\eqno{(4.1c)}\hfill$$
whereas the velocity field defined in a space fixed inertial frame and
projected on the rotating frame
$$\bd{U} = \bd{u} + \bd{\omega}\times\bd{x}\eqno{(4.2)}\hfill$$
A {\em Riemann sequence} is characterized by a parameter 
$$ f = \frac{\zeta}{\omega} - 2\eqno{(4.3)}\hfill$$
When $f=0$, the ellipsoid is rigidly rotating, and when $f=-2$ 
the flow is irotational. Defining the rigidity parameter through the   
relation 
$$ r = 1 + \frac{f}{2}\eqno{(4.4)}\hfill$$
the RR case is reproduced for $r=1$ and the IF case for $r=0$. 
Considering that $\bd{\zeta}$ and ${\omega}$ are parallel with the 
principal axis $x_{3}$, and that $a_{1}\ge a_{2}$, the velocity field 
(4.2) reads 
$$\bd{U}(\bd{r}) = ( 1 - r )\frac{a_{1}^{2}-a_{2}^{2}}{a_{1}^{2}+a_{2}^{2}}
\omega\nabla(x_{1}x_{2}) + r\bd{\omega}\times\bd{r}\eqno{(4.5)}\hfill$$ 
The above equation states that the velocity field $\bd{U(\bd{r})}$ is a 
convex combination of rigid $\bd{U}_{RR} = \bd{\omega}\times\bd{r}$ and 
irotational $\bd{U}_{IF} = 
\frac{a_{1}^{2}-a_{2}^{2}}{a_{1}^{2}+a_{2}^{2}}\omega\nabla(x_{1}x_{2})$ 
contributions. A similar relation is valid for the inertia moment 
${\cal I}_{r}$. A particular example of Riemann flow $r\in (0,1)$ is 
represented in Fig.8(b).

In order to check the predictions given by the 
Riemann model a direct determination of the nuclear current is required.  
The electron-nucleus scattering is an usefull tool which allows the  
measurement of the electromagnetic charge and current densities inside
the ground-state band \cite{20}. 
We showed in a previous section that Coulomb multipoles are associated 
to the charge distribution of the nucleus and thus, in order to obtain 
the quantities depending on the nuclear current we need to determine 
the transversal part of the cross-section, i.e. the electric and magnetic  
multipoles.

Since the task of this work is to substantiate the toroidal 
multipoles which are active in such processes we will foccus  
on the study of electric transverse multipoles arising in the Dubovik 
and Cheshkov multipolar parametrization.

Let us consider an even-even nucleus whose surface oscillates harmonically  
and simultaneously undergo a rigid rotation around an axis perpendicularly 
on its symmetry axis.
Expressing the velocities $\bd{U}_{RR}$ and $\bd{U}_{IF}$ as one and 
two-rank spherical tensors, we readily obtain the spherical component 
$\mu$ of the total velocity.
$$U_{1\mu}(\bd{r}) = (1-r)\left [ V_{2} \otimes r_{1} \right ]_{1\mu} + 
r\left [ V_{1} \otimes r_{1} \right ]_{1\mu}\eqno{(4.6)}\hfill$$ 
where the spherical components of the tensors being coupled to 
$r_{\mu}=\sqrt{4\pi\over 3}rY_{1\mu}$ are given by 
$$V_{1\mu} = -i\sqrt{2}\omega_{\mu},~~~~~V_{2\mu} = 
i\sqrt{{10\over 3}\frac{{\cal I}_{IF}}{{\cal I}_{RR}}}\mu\omega_{\mu}
\eqno{(4.7)}\hfill$$ 
with $\mu = \pm 1$, $\omega_{\mu} = -\frac{\mu\omega}{\sqrt{2}}$, and
${\cal I}_{IF}$ and ${\cal I}_{RR}$ are the inertia moments of IF and  
RR models.

The current density reads
$$\hat{\bd{J}}(\bd{r}) = \rho^{p}(\bd{r}) \bd{U}(\bd{r})\eqno{(4.8)}\hfill$$ 
where
$$\rho^{p} = \sum_{L\geq 2} \rho_{L}^{p}(\bd{r}) Y_{L0}(\theta,\phi)
\eqno{(4.9)}\hfill$$ 
is the proton charge density expanded in even multipolar components 
(~$L=2,4,...$ ) of an axially symmetric nucleus.
 
We shall express the electric transverse (2.4) and longitudinal (2.3)
multipoles as follows \cite{21}
$${\hat T}_{\lambda\mu}^{el}(q) = \frac{i^{\lambda +1}}{\sqrt{2\lambda +1}}
\sum_{\lambda' L} ( \sqrt{\lambda +1}\delta_{\lambda' \lambda -1} -
                    \sqrt{\lambda}   \delta_{\lambda' \lambda +1})\times$$
$$\int_{0}^{\infty} r^{3}dr j_{\lambda'}(qr)\rho_{L}^{p}(r)
\sqrt{3}{\hat L}{\hat \lambda}{\hat \lambda}
\tresj{\lambda'}{1}{L}{0}{0}{0}\seisj{\lambda}{\lambda'}{k}{1}{1}{L} 
(-)^{\mu+k}\tresj{k}{\lambda}{L}{\mu}{-\mu}{-M} V_{k\mu}\eqno{(4.10)}\hfill$$ 
$${\hat L}_{\lambda\mu}^{~~}(q) = \frac{i^{\lambda +1}}{\sqrt{2\lambda +1}}
\sum_{\lambda' L} ( \sqrt{\lambda}   \delta_{\lambda' \lambda +1} +
                    \sqrt{\lambda +1}\delta_{\lambda' \lambda -1})\times$$
$$\int_{0}^{\infty} r^{3}dr j_{\lambda'}(qr)\rho_{L}^{p}(r)
\sqrt{3}{\hat L}{\hat \lambda}{\hat \lambda}
\tresj{\lambda'}{1}{L}{0}{0}{0}\seisj{\lambda}{\lambda'}{k}{1}{1}{L} 
(-)^{\mu+k}\tresj{k}{\lambda}{L}{\mu}{-\mu}{-M} V_{k\mu}\eqno{(4.11)}\hfill$$ 
where $\tresj{.}{.}{.}{.}{.}{.}$ and $\seisj{.}{.}{.}{.}{.}{.}$ are  
$3j$ and $6j$ coefficients \cite{22}.
In the RR case we take $k=1$ and an axial symmetric quadrupole static 
deformation in (4.9) : $\beta=\beta_{2}\neq 0,~\gamma=0$
$$\rho^{p}(\bd{r}) = \frac{3eZ}{4\pi R_{0}^{3}}
\Theta\left [ R_{0}(1 + \beta Y_{20}) -r \right ]\eqno{(4.12)}\hfill$$ 
Thus the charge quadrupole components of the density reads
$$\rho_{2}^{p}(r) = \int d\Omega~Y_{20}^{*}(\theta,\phi)\rho^{p}(\bd{r}) = 
\frac{3eZ\beta}{4\pi R_{0}^{2}}\delta(R_{0}-r)\eqno{(4.13)}\hfill$$ 
and we obtain in place of(4.10) and (4.11)
$${\hat T}_{2\mu}^{el}(q,{\rm RR}) = -Ze \sqrt{3}{\pi}\frac{\sqrt{30}}{40}
\frac{Q_{0}}{R_{0}}\left [ j_{1}(qR_{0}) - {2\over 3}j_{3}(qR_{0}) \right ]
\mu\omega_{\mu}~~\eqno{(4.14)}\hfill$$ 
$${\hat L}_{2\mu}^{~~}(q,{\rm RR}) = -Ze \sqrt{3}{\pi}\frac{\sqrt{30}}{40}
\sqrt{2\over 3}
\frac{Q_{0}}{R_{0}}\left [ j_{1}(qR_{0}) + j_{3}(qR_{0}) \right ]
\mu\omega_{\mu}\eqno{(4.15)}\hfill$$ 
For the irotational case $k=2$ we consider the deformed charge density 
distribution with monopole component
$$\rho_{0}^{p}(r) = \frac{3eZ}{\sqrt{4\pi}R_{0}^{2}}\Theta ( R_{0} - r )
\eqno{(4.16)}\hfill$$ 
such that the multipoles (4.10) and (4.11) may be rewriten
$${\hat T}_{2\mu}^{el}(q,{\rm IF}) = -Ze \sqrt{3}{\pi}\frac{\sqrt{30}}{40}
\frac{Q_{0}}{R_{0}}\left [ j_{1}(qR_{0}) + j_{3}(qR_{0}) \right ]
\mu\omega_{\mu}\eqno{(4.17)}\hfill$$ 
$${\hat L}_{2\mu}^{~~}(q,{\rm IF}) = {\hat L}_{2\mu}^{~~}(q,{\rm RR})\hskip 5.6cm  
\eqno{(4.18)}\hfill$$
Using the convexity property of the velocity field (4.6) the 
electromagnetic multipoles for a certain value of the rigidity 
parameter are
$${\hat T}_{2\mu}^{el}(q,r) = -Ze \sqrt{3}{\pi}\frac{\sqrt{30}}{40}
\frac{Q_{0}}{R_{0}}\left [ j_{1}(qR_{0}) - 
\left ( 1 - {5\over 3}r \right )j_{3}(qR_{0}) \right ] \mu\omega_{\mu}
\eqno{(4.19)}\hfill$$ 
$${\hat L}_{2\mu}^{~~}(q,r) = -Ze \sqrt{2}{\pi}\frac{\sqrt{30}}{40}
\sqrt{2\over 3}
\frac{Q_{0}}{R_{0}}\left [ j_{1}(qR_{0}) + j_{3}(qR_{0}) \right ]
\mu\omega_{\mu}~~~~~~~~~~\eqno{(4.20)}\hfill$$ 
where $Q_{0} = \sqrt{3\over 5\pi}R_{0}^{2}\beta$ is the static quadrupole
moment, and $R_{0}=r_{0}A^{1/3}$. 
The above formulas allow a first interesting remark :
The longitudinal multipole does not depend on the rigidity parameter
and is proportional to the transverse electric multipole
in the IF limit when $r=0$ 
$${\hat L}_{2\mu}(q,r) = \sqrt{2\over 3}{\hat T}_{2\mu}^{el}(q,r=0)
\eqno{(4.21)}\hfill$$ 
Therefore the longitudinal multipoles are insensitive to rotational 
components of the velocity field, their values being constant for any
value of $r$.

Next we will foccus on the behaviour of electromagnetic multipoles at
small momentum transfer.

In 1937, Siegert showed that using the charge-current conservation law
in the longwavelength approximation it is possible to replace the charge  
density operator in the expression of the electric transverse operator
with the charge density rate. Quantitatively this theorem may be expressed 
as follows  
$${\hat T}_{\lambda}^{el}(q\rightarrow 0) = \sqrt{\frac{\lambda +1}{\lambda}}
{\hat L}_{\lambda}(q\rightarrow 0)\eqno{(4.22)}\hfill$$ 
Thus, independently of the model used for the nuclear current the 
transverse electric multipole is proportional to the longitudinal one in 
the longwavelength approximation. 
The approximate equation (4.21) is similar to the exact equation (4.22). 
In other words the Siegert theorem, quantitatively expressed 
by (4.22) is valid in any order of $q$ for the irotational Riemann 
sequence $r=0$. This means that the reactions performed at low-$q$ 
are not able to provide informations on the vorticity.
The Riemann rotator behaves like an irotational liquid drop  
at low momentum transfer. From the view point of the multipolar 
parametrization adopted in this paper this fact may be justified invoking
the following argument : in the low-$q$ limit 
$$\langle I_{f} \parallel {\hat L}_{\lambda}(q\rightarrow 0) \parallel I_{i} 
\rangle = -\frac{iq}{15} \langle I_{f} \parallel 
{\dot {\hat  Q}}_{\lambda}(q\rightarrow 0) \parallel I_{i} \rangle
\eqno{(4.23)}\hfill$$ 
where ${\hat Q}_{2\mu} = \int d\bd{r} r^{\lambda} Y_{\lambda}(\theta,\phi)
{\hat \rho}(\bd{r},t)$ is the charge quadrupole moment. 
Using the charge-current conservation law, the time derivative 
of the charge quadrupole operator may be writen as follows 
$${\dot {\hat Q}}_{2\mu} = \int d\bd{r} r^{\lambda} Y_{\lambda\mu}(\theta,\phi)
\nabla \cdot \hat{\bd{J}}(\bd{r},t)\eqno{(4.24)}\hfill$$ 
The presence of the gradient operator in the above equation ensures the 
cancellation of rotational (vortical) components of the nuclear current.
Thus the time derivative of the charge quadrupole moment describes
the curless quadrupole flows, i.e. $\nabla\times\hat{\bd{J}}=0$, 
like the well known $\beta$ and $\gamma$ vibrations. 
This is the reason why the nuclear response is vibrating-like, without
shear components for small momentum transfer.
Consequently, in order to get informations on the rotational currents 
inside the nucleus, one needs to investigate the electromagnetic 
structures negleted by applying the Siegert theorem. To go beyond the
limitations of this theorem one needs to increase the energy transferred 
in the scattering reaction. The higher order terms in the $q^{2}$ expansion
of the electric transverse multipole are free of the constraint dictated 
by the continuity law and therefore they are likely to provide data on the 
vortical components of the current $\nabla\times\hat{\bd{J}}\neq 0$. 
Applying the Dubovik - Cheshkov parametrization [2] and 
defining a quantity which takes into account the deviation from  
Siegert's theorem
$$\eta_{2}(q) = q^{2} \frac
{\langle I_{f} \parallel {\hat T}_{2}^{tor}(q) \parallel I_{i}\rangle}
{\langle I_{f} \parallel {\hat T}_{2}^{el~} (0) \parallel I_{i}\rangle}
\eqno{(4.25)}\hfill$$
we are in the position to describe the importance of the toroidal 
multipoles \cite{23}.
In the IF and RR cases, the function $\eta_{2}$ looks 
$$\eta_{2}(q,{\rm IF}) = \frac{15}{(qR_{0}^{2})^{5}}
\left [ ( 3 - (qR_{0})^{2})\sin qR_{0} - 3qR_{0}\cos qR_{0} \right ] -1\hskip 2.85cm
\eqno{(4.26)}\hfill$$
$$\eta_{2}(q,{\rm RR}) = \frac{15}{(qR_{0}^{2})^{6}}
\left [ \left ( (qR_{0})^{2} - 2 \right )\sin qR_{0} + 
\left ( 2 - {(qR_{0})^{2}\over 3} \right ) qR_{0}\cos qR_{0} \right ] -1
\eqno{(4.27)}\hfill$$
whereas in the intermediate case, this function will be expressed as a  
convex combination of IF and RR contributions 
$$\eta_{2}(q,r) =  q^{2} \frac
{\langle I_{f} \parallel {\hat T}_{2}^{tor}(q,r) \parallel I_{i}\rangle}
{\langle I_{f} \parallel {\hat T}_{2}^{el} (0) \parallel I_{i}\rangle} = 
( 1- r )\eta_{2}(q,r=0) + r\eta_{2}(q,r=1)
\eqno{(4.28)}\hfill$$
In Fig.9 we draw this quantity versus momentum transfer $q$. 
Notice that the RR model presents a stronger deviation from the Siegert
approximation than the IF model. 
In the hexadecupole case ( $\lambda=4$ ) the deviation effect is smaller
than in the quadrupoe case ( $\lambda=2$ ) for both models.

Another interesting quantity is the real electric transverse form factor                
which in the lowest order of $q^{2}$ is
$${\cal F}_{\lambda}^{el}(q) \equiv \frac 
{\langle I_{f} \parallel {\hat T}_{\lambda}^{el}(q) \parallel I_{i}\rangle}
{\langle I_{f} \parallel {\hat T}_{\lambda}^{el} (0) \parallel I_{i}\rangle} 
\approx 1 - \frac{q^{2}}{3} {\cal I}_{r}\frac{{\cal T}_{2}}{Q_{2}}
\eqno{(4.29)}\hfill$$
where $Q_{2}=3e^{2}ZR_{0}^{2}\beta/4\pi$ is the transition charge 
quadrupole moment, and ${\cal T}_{2}$ is the transition toroidal 
quadrupole moment. This last equation allows the measurement of 
the transition toroidal quadrupole moment ${\cal T}_{2}$ similar to 
that of the charge (magnetic) mean square radius $< r^{2} >_{C(mag)}$ [24],  
which consists in the computation of ${\cal F}_{2}^{el}(q)$. 
For an arbitrary value of the rigidity parameter the toroidal quadrupole 
moment reads \cite{25}
$${\cal T}_{2}(r) = 
\frac{3e^{2}Z}{56\pi}\frac{3+2r}{{\cal I}_{r}} R_{0}^{4}\beta
\eqno{(4.30)}\hfill$$
\vspace{5mm}

We ploted in Fig.10 ${\cal T}_{2}$ versus the rigidity for $^{152}$Sm and 
$^{166}$Er. An important conclusion that we draw from this figure is
is that ${\cal T}_{2}$ increases sharply for values of $r$ close to 1, 
being two orders of magnitude larger in the RR case than in the IF case. 
We conclude that in those nuclei where the vortical components of the  
nuclear current are stronger the toroidal quadrupole transitions are  
intensified with respect to the spherical vibrating nuclei.

Taking into account the definition of the toroidal multipoles, their
measurement is equivalent with the measurement of electric transverse 
multipole at high momentum transfer followed by the removal of the  
Siegert limit.

In the excitations of ground state band of an even-even nucleus
there will be involved the electric multipoles with $\lambda = 2,4,...$ 
and and the magnetic multipoles with $\lambda = 1,3,...$. 
Further we will not discuss the magnetic multipoles and we will foccus
on the longitudinal and electric transverse parts 
of the differential cross-sections.
Since our purpose is to shed light onto the nature of toroidal transitions
and their connection with nuclear vortical currents we will chose a 
convenient method to separate the dominant longitudinal components in the 
differential cross section and eventually to obtain the transverse 
multipoles.
Separating the transverse multipoles by 180$^{0}$ scattering is a 
seducing method since at this angle these multipoles dominate 
the differential cross section.
Another method of separation consisting in the scattering of polarized  
electrons will be discussed bellow. 

For the $0^{+}\rightarrow\lambda^{+}$ excitation we have the following 
expression for the differential cross-section:
$$\left ( \frac{d \sigma}{d \Omega} \right )_{(e,e')}  = 
\frac{4\pi\sigma_{Mott}}{f_{rec}}
\left \{ \frac{q_{\mu}^4}{\bd{q}^4} 
|\langle\lambda^{+}\parallel {\hat M}_{\lambda}(q)\parallel 0^{+}\rangle|^{2} 
+ \left (\frac{q_{\mu}^{2}}{2q^{2}}+\tan^{2}{\theta\over 2} \right )
|\langle \lambda \parallel {\hat T_{\lambda}^{el}}(q)\parallel 0^{+} 
\rangle|^{2} \right\}\eqno{(4.31)}\hfill$$
Using the Dubovik - Cheshkov decomposition in the above formula and  
neglecting the Coulomb multipoles and the low-$q$ limit of electric 
transverse multipole one get
$$\left ( \frac{d \sigma}{d \Omega} \right )_{(e,e')} = 
\frac{4\pi\sigma_{Mott}}{f_{rec}} q^{2}
\left (\frac{q_{\mu}^{2}}{2q^{2}}+\tan^{2}{\theta\over 2}\right )
|\langle \lambda \parallel {\hat T_{\lambda}^{tor}}(q)\parallel 0^{+} 
\rangle|^{2} \eqno{(4.32)}\hfill$$
This approximation is equivalent to the neglecting of transition 
charge moments $Q_{\lambda}$ mean 2$n$-power of charge distribution 
radii $r_{\lambda}^{2n}$.

The differential cross sections (4.31) and (4.32) are ploted in both  
cases, for RR and IF, in order to compare the exact formula and the  
approximate one in backscattering processes.
The cross section of the quadrupole transition induced by the scattered 
electron is represented in Fig.11 for the $^{152}$Sm nucleus. 
The hexadecupolar transition is considered in Fig.12.
From the study of these two graphics it becomes obvious that 
the above mentioned approximation is acceptable in the RR model 
for the whole range of momentum transfer for both considered transitions.
The main difference consists in the location of diffraction minima. 
These differences are not important because in a phase-shift analysis 
the curve will be smothed in the neighborhood of minimas.

In the IF case, the exact differential cross section (4.31) matches quite 
well the approximate one for momentum transfer $q < $400 MeV$/c$ 
when $\lambda = 2$ and $q < $250 MeV$/c$ when $\lambda$ = 4 
for the case $^{166}$Er. 
The reason of these discrepancies in the IF model between the 
exact and approximate curves is that at high momentum transfer 
the mean 2$n$-power charge distribution radius is drastically enhanced. 

\section{Separation of toroidal multipoles in electron coincidence
processes}
As we saw in the preceding sections of this work a method to separate 
the electric transverse multipoles is given by the Rosenbluth decomposition  
of the $(e,e')$ differential cross section at $180^{0}$ scattering 
angle. A more recent method which allows this separation is based on the 
$(e, e'\gamma)$ longitudinal/transversal interferences \cite{26}. 
The differential cross section of the process is given by         
$$
\left ( \frac{d^{2}\sigma}{d\Omega_{e}d\Omega_{\gamma}}\right )^{h,\sigma} =  
{1\over 2}\Sigma_{0}^{\lambda^{+}0^{+}}
\left (\frac{\Gamma_{\gamma}^{\lambda^{+}\rightarrow (\lambda -2)^{+}}}
{\Gamma_{total}^{\lambda^{+}}}\right )
\left ( W_{\Sigma}(\theta_{\gamma},\phi_{\gamma}) + 
h\sigma W_{\Delta}(\theta_{\gamma},\phi_{\gamma})\right ) \eqno{(5.1)}\hfill$$
where $\Sigma_{0}^{\lambda^{+}0^{+}}$ is the differential cross section of  
the corresponding $(e, e')$ process, given by (4.31), 
$\Gamma_{\gamma}^{\lambda^{+}\rightarrow (\lambda -2)^{+}}$ is the 
photodisintegration width of the transition 
$\lambda^{+}\rightarrow (\lambda -2)^{+}$ and 
$\Gamma_{total}^{{\lambda}^{+}}$ is the total decay width for the state
$\lambda^{+}$. The ratio of these two widths is close to unity. 
The differential cross section contains as labels the incident electron 
helicity, $h = \pm 1$, and the circular polarization of the photon
detected in coincidence, $\sigma = \pm 1$. The angular distribution 
functions are normalized to unity ( $\int d\Omega_{\gamma}W_{\Sigma} = 1$ ), 
and satisfy the integral condition 
( $\int d\Omega_{\gamma}W_{\Delta} = 0$ ). Their explicite expressions are
$$W_{\Sigma}(\theta_{\gamma},\phi_{\gamma}) = 
{1\over 4\pi}\left ( 1 + A_{\Sigma}(\theta_{\gamma},\phi_{\gamma})\right )
\eqno{(5.2)}\hfill$$
$$W_{\Delta}(\theta_{\gamma},\phi_{\gamma}) = 
{1\over 4\pi} A_{\Delta}(\theta_{\gamma},\phi_{\gamma})\hskip 1.1cm\eqno{(5.3)}\hfill$$

In the electroexcitation of the rotational g.s. band of a deformed 
even-even nucleus ( 0$^{+}$, 2$^{+}$, 4$^{+}$,...), the longitudinal 
multipoles ( C2, C4,... ) are larger than the electric transvere 
multipoles ( E2, E4,... ). 
The even-even nuclei are not good candidates to become polarized targets,  
because they have spin and parity 0$^{+}$ in the ground state. 
However they are good candidates for the $(e, e'\gamma)$ study 
where the L/T interferences can be isolated and the matrix
elements of the electric transverse multipole computed.
Introducing the ratio between the electric transverse multipole and  
the Coulomb one 
$$\xi_{\lambda} = \frac{\langle I_{f}\parallel {\hat T}_{\lambda}^{el}\parallel I_{i}\rangle}
     {\langle I_{f}\parallel {\hat M}_{\lambda}^{~}\parallel I_{i}\rangle}
\eqno{(5.4)}\hfill$$
and considering that $\xi_{\lambda}\ll 1$, there will be considered 
only the linear terms in the angular distribution function. 
The transversal-longitudinal interference term, which is linear  
in $\xi_{\lambda}$ is the most interesting. It can be isolated performing 
measurements at $\phi_{\gamma} = 0^{o}$ and $\phi_{\gamma} = 180^{0}$ with 
unpolarized electrons
$$
\left.\left.
\frac{d^{2}\sigma}{d\Omega_{e} d\Omega_{\gamma}}
\right|_{\phi_{\gamma}=0^{o}}^{nepol} -
\left.
\frac{d^{2}\sigma}{d\Omega_{e}d\Omega_{\gamma}}
\right|_{\phi_{\gamma}=180^{0}}^{nepol}
\right /\left( 
\frac{\Gamma_{\gamma}^{\lambda^{+}\rightarrow (\lambda -2)^{+}}}
{\Gamma_{total}^{\lambda^{+}}}\right )4\pi\sigma_{Mott}f_{rec}^{-1} = $$
$$-\sqrt{2}\frac{q_{\mu}^{2}}{q^{2}}\sqrt{\frac{q_{\mu}^{2}}{q^{2}} + 
\tan^{2}{\theta\over 2}} G_{TL}\xi_{\lambda}
\left ( \langle I_{f} \parallel {\hat M}_{\lambda}(q)\parallel I_{i}\rangle
\right )^{2}\eqno{(5.5)}\hfill$$
where 
$$G_{TL} = \frac{\sqrt{2}}{7(2\lambda -1)}\sqrt{\frac{\lambda +1}{\lambda}}
\left \{ 5P_{2}^{1}(\cos\theta_{\gamma}) - 
3\left (\frac{\lambda +2}{2\lambda-3}\right )
P_{4}^{1}(\cos\theta_{\gamma})\right \}\eqno{(5.6)}$$
Therefore, chosing convenient values which maximize (5.5), it is possible
to determine
$\langle I_{f} \parallel {\hat M}_{\lambda}(q)\parallel I_{i}\rangle 
\langle I_{f} \parallel {\hat T}_{\lambda}^{el}(q)\parallel I_{i}\rangle$ 
for different values of $q$.

Let us consider as an example the study of $(e, e'\gamma)$ processes  
when the excited states are vibrational collective modes. 
In the incompressible liquid drop model the density operator is given
by
$${\hat \rho}_{N}(\bd{r}) = \frac{3eZ}{4\pi R_{0}^{3}}
\Theta\left [ R_{0}(1 + \sum_{lm}\alpha_{lm} Y_{lm}^{*}) -r \right ]
\eqno{(5.7)}$$
and the current density operator by
$$\hat{\bd{J}}_{N}(\bd{r}) = 
\frac{3eZ}{4\pi R_{0}^{3}}\sum_{lm}{1\over l}
\dot\alpha_{lm}\left [ \nabla\left ({r\over R_{0}} \right )^{l} 
Y_{lm}(\theta,\phi) \right ] \Theta(R_{0} - r)
\eqno{(5.8)}\hfill$$
For the transition to the one-surfon 2$^{+}$ state, the Coulomb multipole 
is
$$\langle 2^{+} \parallel {\hat M}_{2}(q)\parallel 0^{+}\rangle = 
\frac{3eZ}{4\pi}\sqrt{\frac{5}{2(B_{2}C_{2})^{1/2}}}j_{2}(qR_{0})
\eqno{(5.9)}\hfill$$
where $B_{2}$ and $C_{2}$ are the inertia and rigid parameters.
Since the liquid drop velocity field is postulated to be irotational, 
the contribution of the transverse multipole reads 
$$\langle 2^{+} \parallel {\hat T}_{2}^{el}(q)\parallel 0^{+}\rangle = 
-\frac{\omega_{2}}{q}\sqrt{3\over 2}
\langle 2^{+} \parallel {\hat M}_{2}(q)\parallel 0^{+}\rangle 
\eqno{(5.10)}\hfill$$
with $\omega_{2} \approx 36 A^{-1/2} {\rm MeV}$. 
In order to take into account the magnetization components of the current
in eq.(5.8), a crude model for the magnetization density is used 
\cite{14}
$${\hat \mu}_{N}(\bd{r}) = \frac{\mu}{2mZ}{\hat \rho}_{N}(\bd{r})\bd{L}
\eqno{(5.11)}\hfill$$
and the electric transverse multipole becomes
$$\langle 2^{+} \parallel {\hat T}_{2}^{el}(q)\parallel 0^{+}\rangle = 
-\frac{\omega_{2}}{q}\sqrt{3\over 2}
\left ( 1 -\frac{\mu}{Z}\frac{q^{2}}{2m\omega_{2}}\right )
\langle 2^{+} \parallel {\hat M}_{2}(q)\parallel 0^{+}\rangle 
\eqno{(5.12)}\hfill$$
Introducing the deviation from the Siegert theorem (4.25) it is then 
possible to establish a connection with the interference factor  
(5.4) as follows \cite{23}
$$\eta_{2}(q) = 1 + \sqrt{2\over 3}\frac{q}{\omega_{2}}\xi_{2}(q)
\frac{\langle 2^{+} \parallel {\hat M}_{2}(q)\parallel 0^{+}\rangle}
{\langle 2^{+} \parallel {\hat M}_{0}(q)\parallel 0^{+}\rangle}
\eqno{(5.13)}\hfill$$
The dependence of $\eta_{2}$ on momentum transfer $q$
is given in Fig.13, for $^{16}$O and $^{90}$Zr in both cases : 
irotational flow and with nonzero magnetization ( $\mu = 0.5$ ). 
Notice that for $^{16}$O the magnetization contributions have a  
sensitive effect : the deviation from the Siegert theorem is enhanced. 
For the nucleus $^{90}$Zr, the magnetization contributions are less  
important.

As we saw earlier the deviation from the Siegert theorem increases when 
we pass from the IF model to the RR model. Although the microscopic 
nature of the magnetization current is different from that of the RR current
from geometrical point of view they have the same rotational structure 
leading to the same effect : enhancement of the toroidal transitions. 
It can be also noticed that the deviation from the Siegert theorem is
stronger in heavy nuclei than in light nuclei. This last conclusion  
is in agreement with the statement that we have made on the electric 
transverse form factor of DTM, i.e. the toroidal effect is intensified in
heavy nuclei.

\section{Conclusions}
As we mentioned in the begining the scope of this work was the study  
of collective motions with a toroidal electromagnetic structure or to
extract the toroidal contribution in nuclear transitions with a 
mixed rotational-vibrational spectrum.

Considering the giant isoscalar resonance 1$^{-}$ DTM, we calculated 
for a group of spherical nuclei ( $^{40}$Ca, $^{90}$Zr and $^{208}$Pb )
the transverse form factors and we underlined the shift of the toroidal
effect toward small momentum transfers when we pass to nuclei with 
large $Z$ and $A$. Simultaneoulsy we testify an enhancement of the  
dipole toroidal response in heavy nuclei.
The probability to excite the DTM with photon probes is small 
according to the calculation we have made. However the possibility  
to use electron inelastic scattering offers the promise to detect DTM
because in such reactions it is possible to vary independently the   
momentum transfer and the excitation energy along with the scattering angle 
$\theta$ dependence of the differential cross section. Since the    
DTM is a transversal flow associated to the Hill ring vortex the  
electroexcitation at 180$^{0}$ angles seems to be favourable since
other modes which could occur, being especially of longitudinal nature,  
are suppressed. We have also extracted the longwavelength limit of the 
electric transverse form factor and we emphasized that it is proportional 
with the transition toroidal dipole moment. This dynamic characteristic 
associated to the DTM for small momentum transfers depends linearly on
$\alpha Z$, which explains the smallness of toroidal effects for   
nuclei with small numbers of protons.

Another type of resonances that we studied are the spin dependent (spin - 
flip) modes. We emphasized their purely toroidal electromagnetic structure   
and their possible investigation with leptonic probes at 180$^{0}$ 
scattering angles and large momentum transfer. These conclusions stems on 
the calculation of $(e, e')$ differential cross sections of 
Goldhaber - Teller resonances and  1$^{-}$ electric spin - flip modes 
(s-is) in $^{12}$C, $^{16}$O, $^{40}$Ca and $^{208}$Pb.

In the second part of this work we foccused on the study of toroidal
contributions in the excitation of g.s. band of even-even nuclei 
from the rare-earth region (~$^{152}$Sm, $^{166}$Er ). 
We introduced a quantity which describes the deviations from the Siegert 
theorem or in other terms the contribution of higher order terms in the 
momentum transfer of the electric transverse multipole, i.e. the toroidal 
multipole moments and their mean square radii.
For the RR which is a submodel of the Riemann rotator the deviation from
the Siegert theorem has a slope larger than that of the IF model.  
Based on this observation we showed that defining the real electric 
transverse form factor and taking the first order approximation 
in the momentum transfer we can extract the transition toroidal 
quadrupole moment in the same manner in which we determine the charge or 
magnetic mean square radii : calculating the slope of the real Coulomb 
(magnetic) form factor. 

We computed the transition toroidal quadrupole moment and we determined 
that it depends smoothly on the vorticity
This observation strengths our opinion that the toroidal moments of  
arbitrary multipolarity give a measure of the intensity of vortical
electromagnetic currents in the same maner the charge multipole moments  
are associated to charge distribution or irotational electromagnetic 
currents.

The electroexcitation differential cross sections of 2$^{+}$ and 4$^{+}$ 
levels from the g.s. band of the RR model may be approximated taking  
into account taking into account the toroidal form factors only, for a
wide range of momentum transfer.
This result emphasize the importance of toroidal multipoles relative to  
the Coulomb multipoles at 180$^{0}$ angles, regardless of the momentum 
transfer.

We showed that in coincidence $(e, e'\gamma)$ reactions, by separating  
the longitudinal/transversal interference term we can measure the 
deviation from the Siegert theorem in the IF model and with non-zero 
magnetization components of the nuclear current.
For light nuclei the deviation from the Siegert theorem is enhanced 
by the existence of magnetic currents.

\newpage


\newpage
\begin{thebibliography}{99}
\bibitem{1} V.M.Dubovik and A.A.Ceshkov, Fiz.El.Chast.At.Yadra 
{\bf 5} (1974) 791
\bibitem{2} V.M.Dubovik and V.V.Tugushev, Phys.Rep. {\bf 187}, 
no.4 (1990) 145
\bibitem{3} Ya.B.Zeldovici, JETF {\bf 33} (1957) 1531
\bibitem{4} V.V.Flambaum and I.B.Khriplovici, 
JETF {\bf 79} (1980) 1656
\bibitem{5} \c S.Mi\c sicu, Rom.J.Phys. {\bf 37}, no.9 (1992) 855
\bibitem{6} G.N.Afanasiev, V.M.Dubovik and \c S.Mi\c sicu, 
J.Phys.A : Math.Gen. {\bf 25} (1992) 4869
\bibitem{7} G.F.Bertsch, Nucl.Phys. {\bf A249}(1975) 253
\bibitem{8} P.M.Morse and H.Feschbach, {\em Methods of 
Theoretical Physics}, McGraw - Hill, New York, 1952 
\bibitem{9} S.I.Bastrukov, \c S.Mi\c sicu and 
A.V.Sushkov, Nucl.Phys. {\bf A562} (1993) 191
\bibitem{10} L.M.Milne-Thompson, {\it Theoretical Hydrodynamics}, 
ed. 5-a, Macmillan, Londra, 1968
\bibitem{11} S.F.Semenko, Yad.Fiz. {\bf 34} (1981) 639 
\bibitem{12} E.B.Balbutsev and I.N.Mikhailov, 
J.Phys.G:Nucl.Part.Phys. {\bf 14}(1988) 545
\bibitem{13} J.R.Nix and A.J.Sierk, Phys.Rev. {\bf C21}, 
no.1(1980) 396
\bibitem{14} 
T.de Forest and J.D.Walecka, Adv.Phys. {\bf 15}, no.57(1966) 1
\bibitem{15} 
J.Heisenberg \c si H.P.Block, Ann.Rev.Nucl.Part.Sci. {\bf 33} (1983) 569
\bibitem{16}  D.G.Ravenhall and J.Wambach, Nucl.Phys. {\bf A475} 
(1987) 468
\bibitem{17} V.M.Dubovik and L.A.Tosunyan, Fiz.El.Ciast.At.Yadra, 
{\bf 14}, n.5 (1983) 1193
\bibitem{18} E.P.Wigner, Phys.Rev. {\bf 51} (1937) 106
\bibitem{19} G.Rosensteel, Ann.Phys. {\bf 186} (1988)230
\bibitem{20} E.Moya de Guerra, Phys.Rep. {\bf 138}, 
no.6 (1986) 293
\bibitem{21} G.Rosensteel, Phys.Rev. C{\bf 41}, no.3 (1990) R811
\bibitem{22} A.R.Edmonds, {\it Angular Momentum in Quantum 
Mechanics}, Princeton University Press, Princeton, New Jersey, 1960
\bibitem{23} \c S.Mi\c sicu, in {\em Frontier Topics in 
Nuclear Physics}, p.262, ed.W.Scheid and \\ A.S\u andulescu, Plenum Press,  
New York, 1994
\bibitem{24}  H.Uberall, {\it Electron Scattering from Complex 
Nuclei, PartA}, Academic Press, N.Y., 1971
\bibitem{25} \c S.Mi\c sicu, 
J.Phys.G:Nucl.Part.Phys. {\bf 21} (1995) 669
\bibitem{26} C.Garcia-Recio, T.W.Donnelly and E.Moya de Guerra,  
Nucl.Phys. {\bf A509} (1990) 221
\bibitem{27} \c S.Mi\c sicu, in {\em Selected Topics on 
Nuclear Structure, Dubna, 5-9 July}, pag.77, ed.V.G.Soloviov, K.Ya.Gromov 
\c si L.A.Malov, vol.1(1994).   
\end{thebibliography}
\end{document}